# Monte-Carlo-based electromagnetic modeling of nanoscale structures: accounting for inhomogeneous broadening in polydisperse ensembles

Herman Gudjonson[1*], Mikhail A. Kats[1*], Kun Liu[2,3], Zhihong Nie[4], Eugenia Kumacheva[3,5,6], and Federico Capasso[1+]

[1]*School of Engineering and Applied Sciences, Harvard University, Cambridge, Massachusetts 02138, USA*

[2]*State Key Laboratory of Supramolecular Structure and Materials, College of Chemistry, Jilin University, Changchun 130012, P. R. China*

[3]*Department of Chemistry, University of Toronto, 80 St. George Street, Toronto, Ontario M5S 3H6, Canada*

[4]*Department of Chemistry and Biochemistry, University of Maryland, College Park, MD 20742*

[5] *Department of Chemical Engineering and Applied Chemistry, University of Toronto, 200 College Street, Toronto, Ontario M5S 3E5, Canada*

[6] *The Institute of Biomaterials and Biomedical Engineering, University of Toronto, 4 Taddle Creek Road, Toronto, Ontario M5S 3G9, Canada*

*Equal contribution

[+]Corresponding author: capasso@seas.harvard.edu

**Abstract:**

Many experimental systems consist of large ensembles of uncoupled or weakly interacting elements operating as a single whole; this is the case in many experimental systems in nano-optics and plasmonics including colloidal solutions, plasmonic nanoparticles, dielectric resonators, antenna arrays, and others. In such experiments, measurements of the optical spectra of ensembles will differ from measurements of the independent elements even if these elements are designed to be identical as a result of small variations from element to element, known as polydispersity. In particular, sharp spectral features arising from narrow-band resonances will tend to appear broader and can even be washed out completely. Here, we explore this effect of inhomogeneous broadening as it occurs in colloidal nano-polymers comprising self-assembled nanorod chains in solution. Using a technique combining finite-difference time-domain (FDTD) simulations and Monte-Carlo sampling, we predict the inhomogeneously-broadened optical spectra of these colloidal nano-polymers, and observe significant qualitative differences compared to the unbroadened spectra. The approach combining an electromagnetic simulation technique with Monte-Carlo sampling is widely applicable for quantifying the effects of inhomogeneous broadening in a variety of physical systems, including those with many degrees of freedom which are otherwise computationally intractable.

**Introduction**

In photonics experiments and applications, frequent use is made of ensembles of individual structures operating as a single whole; these include, for example, lithographically-defined arrays of metallic nanostructures which form frequency selective surfaces [1], metasurfaces [2] [3] [4] or sensor arrays [5], colloidal solutions or suspensions [6] [7], randomly dispersed nanoshells, quantum dots or nanocrystals on a substrate [8], and many others.

Assuming that the elements in the ensembles are independent (i.e. they do not experience significant near- or far-field coupling), an assumption that can often be made in sparse, disordered systems, the optical response of these ensembles is simply the sum of the response of all of their constituents. In the case that such an ensemble is composed of many identical elements, its spectral response should be the same as that of each individual element. In real systems, however, the constituent elements are never precisely identical: any fabrication or synthesis technique including top-down lithography and bottom-up self-assembly will introduce a distribution of geometrical and parameters (a.k.a. polydispersity) which leads to inhomogeneous broadening in the spectral features of the total ensemble (e.g. [9] [10] [11] [12] [13] [14] [15] [16]). To avoid inhomogeneous broadening effects in experiments, complex techniques are sometimes used to measure the optical response of individual elements [17]. Other times inhomogeneous broadening can be helpful, for example in situations where a broadband optical response is desired such as in photovoltaic applications [18].

While full-wave electromagnetic simulations are often used to model and understand optical systems that cannot be described analytically (e. g. [19]), these methods cannot easily account for polydispersity which leads to inhomogeneous broadening. This issue is sometimes addressed by artificially increasing the damping constant of materials [15], but this approach only yields loose, qualitative information, does not provide a way to distinguish between the various sources of polydispersity (geometrical or material), and is in general not physical.

In the present work, we demonstrate that a complex ensemble of non-interacting elements can be fully modeled using a Monte-Carlo approach [20] [21], utilizing finite-difference time-domain (FDTD) simulations for the intermediate steps. Monte-Carlo methods combined with electromagnetic calculations have previously been applied to problems in electromagnetics such as scattering from random rough surfaces [21] [22] [23] [24] and light transport through tissues [25], but to our knowledge have not been utilized to study the effects of inhomogeneous broadening in photonic or plasmonic ensembles. Here we predict the extinction spectra of self-assembled gold nanorod chains ("nano-polymers") suspended in a solution. This physical system has a large number of degrees of freedom (e.g. the lengths and widths of the individual nanorods comprising the chains, the total number of rods comprising each chain, the gaps between the rods, their orientation, etc), and is therefore a particularly challenging model system.

**Model system: gold nanopolymers in solution**

Recent experiments have demonstrated that gold nanorods end-tethered with polystyrene ligands can undergo self-assembly in solution and form linear (or bent) chains, in a process analogous to step-growth polymerization [26] [27] [28]. In this process, individual nanorods with an end-tether on each end behave as monomers (Fig. 1(a)). In a colloidal polymer (a "nanopolymer"), the nanorods are the repeat units and the tethers between the nanorod ends act as bonds.

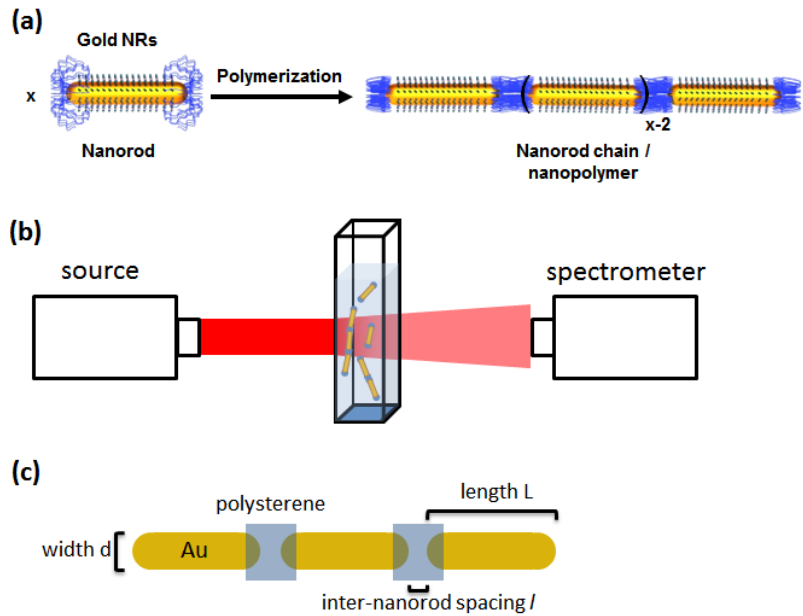

Figure 1. Self-assembled nanorod (NR) chains. (a) Self-assembly of gold nanorods end-tethered with polystyrene functional groups to form nanorod chains. (b) Hypothetical experimental setup: broadband light source incident on a nanorod solution: a spectrometer records light that is not absorbed or scattered by the solution (unity minus extinction). (c) nanorod chain model with constant geometrical parameters $L$, $d$, and $l$. The background and polymer refractive indices are $n_{background}$ = 1.42 and $n_{dielectric}$ = 1.57, respectively.

At a particular stage of self-assembly, the concentration of unreacted functional groups $[M]$ is twice as large as the concentration of nanorod chains in the system (including individual unreacted nanorods), since each chain has two ends. The number average degree of polymerization $\overline{X}_n$ is defined as

$$\overline{X}_n = \frac{\text{total number of NRs}}{\text{total number of chains}} = \frac{[NR]_0}{\frac{1}{2}[M]} \qquad \text{(Eq. 1)},$$

where $[NR]_0$ is the initial concentration of nanorods in the solution. For the self-assembly time $t = 0$, $\overline{X}_n = 1$, and the initial concentration of functional groups is $[M]_0 = 2[NR]_0$.

The self-assembly occurs as follows: the first step is the reaction between two individual nanorods to form a dimer; the dimer can then react with a monomer to form a trimer or with another dimer to form a tetramer, and so forth [28]. This process yields a mixture of chains comprising various numbers of nanorods $x$. The degree of polymerization of the entire mix of nanorods and nanorod chains at a particular moment in time can be quantified by the "conversion" parameter $p$, defined as the fraction of end-tethers that have reacted, such that

$$[M] = [M]_0 - [M]_0 p = [M]_0 (1 - p) \qquad \text{(Eq. 2)}$$

The conversion $p$ is then related to $\overline{X}_n$ by

$$\overline{X}_n = \frac{[NR]_0}{\frac{1}{2}[M]} = \frac{[M]_0}{[M]} = \frac{1}{1-p} \qquad \text{(Eq. 3)}.$$

For this type of step-growth polymerization, the concentration of chains containing $x$ nanorods, $c_{x,p}$, can be predicted by the Flory (or "most probable") distribution given a particular conversion $p$ as [28] [29]

$$c_{x,p} = [NR]_0 (1-p)^2 p^{(x-1)} \qquad \text{(Eq. 4)}.$$

A solution of gold nanorods and nanorod chains ($x$-mers) can be viewed as a "metamaterial fluid" or "metafluid", and will have different optical properties from that of the solvent alone as a result of the resonant scattering contribution of the $x$-mers. In the language of metafluids, the $x$-mers can be viewed as "artificial plasmonic molecules" [7] suspended in a liquid, and one can envision a characterization experiment in which the extinction spectrum of the fluid is measured using a broadband optical source and a spectrometer (Fig. 1(b)). If the solution is dilute and there is little agglomeration of $x$-mers, the extinction spectrum of the solution can be calculated as the sum of the extinction spectra of all of the individual $x$-mers, which can be predicted by a variety of full-wave electromagnetic simulation techniques [30]. Since the solution is dilute, any multiple scattering effects can be neglected, and the spectra can be summed incoherently (i.e. neglecting phases) because the positions and orientations the individual nanorod chains are random and constantly changing throughout the solution via thermal motion. Nonetheless, the problem remains very challenging because of the large number of degrees of freedom: a solution can contain chains of nanorods of nearly any length, and every nanorod can differ in its geometrical parameters.

**Modeling nanorod chains comprising identical (monodisperse) nanorods**

The extinction spectrum of a particular nanorod chain is determined by the length ($L$) and diameter ($d$) of the individual nanorods comprising the chain, the inter-nanorod distance ($l$), and the number $x$ of nanorods in the chain (see, e.g., [31]). For simplicity, we assumed that the chains remain linear (no bending). To model the spectrum of an individual nanorod chain, we first assumed that the values of $L$, $d$, and $l$ are constant for all nanorods comprising the chain, and then examined the relationship between $x$ and the normalized extinction spectrum $\varepsilon_x(\lambda)$ of an individual chain with a particular, well-defined aggregation number $x$, as $x$ increased from 1 to 10. $\varepsilon_x(\lambda)$ is determined as $\varepsilon_x(\lambda) = \sigma_x(\lambda)/x$ where $\sigma_x(\lambda)$ is the extinction cross section of the chain, and represents the extinction spectrum of a single chain normalized to its length.

We performed full-wave three-dimensional FDTD simulations using the total-field scattered-field (TFSF) formulation [32], implemented in a commercial software package (FDTD Solutions). In the simulations, we used $L = 52$ nm, $d = 13$ nm, and $l = 6.7$ nm, the values corresponding to those in self-assembly experiments. The mesh size of the simulations was 0.5 nm such that all features were well resolved. We used a background index of refraction of ~1.42 corresponding to typical solvents used in nanorod self-assembly (N,N-dimethylformamide (DMF) - water mixture with a water content of 15 wt% [33]) and an index of refraction of ~1.57 for the polystyrene ligands [34]. The incident light was set to be polarized with the electric field along the long axis of the chain as that is the orientation of the dominant dipole moment of the chains.

The simulated normalized extinction spectra $\varepsilon_x(\lambda)$ of these nanorod chains with $x$ from 1 to 10 are plotted in Fig. 2(a). As $x$ increased, the localized surface plasmon resonance (LSPR) peak experienced a gradual red-shift from ~890 nm to ~1020 nm. This type of red-shift has been consistently predicted and observed in plasmonic dimers, trimers, and longer chains and is generally attributed to a combination of capacitive near-field coupling between the neighboring nanorods and retardation effects which set in when the size of the chain becomes non-negligible compared to the wavelength [31] [35] [36] [37] [38]. The effect was particularly strong as $x$ increased from 1 to 4, as the majority of nanorods forming the chain acquired new nearest neighbors, but then quickly saturated for longer chains.

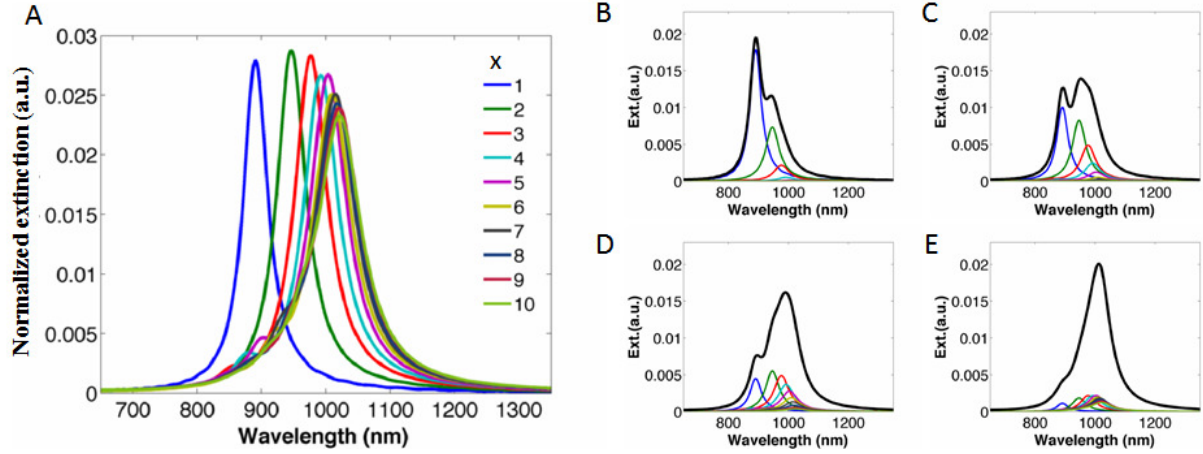

Figure 2. Simulated extinction spectra of chains comprising nanorods of constant size. (a) Normalized extinction spectrum $\varepsilon_x(\lambda) = \sigma_x(\lambda)/x$ of individual chains comprising $x$ nanorods, each with $L = 52$ nm, $d = 13$ nm, and $l = 6.7$ nm. (b-e) Simulated extinction spectra of Flory-distributed ensembles of monodisperse nanorod chains with conversions $p$ of 0.3, 0.5, 0.7, and 0.9, respectively. The thin colored lines show the contribution to the overall extinction spectrum from the chains with a particular $x$, and the thick black lines show the total extinction spectrum of the entire population of chains ($Ext_{tot,p}(\lambda)$).

Next, we modeled the extinction spectra of a population of monodisperse nanorod chains at various conversions $p$. For dilute colloidal solutions, the total extinction, $Ext_{tot,p}$, is the sum of extinctions of the individual species. Thus for non-interacting, monodisperse nanorod chains

$$Ext_{tot,p}(\lambda) = b(\sigma_1(\lambda)c_{1,p} + \sigma_2(\lambda)c_{2,p} + \sigma_3(\lambda)c_{3,p} + ...) = b\sum_{x=1}\sigma_x(\lambda)c_{x,p} \qquad (Eq.\ 5),$$

where $b$ is the interaction path length, $\sigma_x(\lambda)$ is the extinction cross-section of an individual nanorod chain, $c_{x,p}$ is the concentration of the nanorod chains with a particular $x$ at conversion $p$, and the subscripts 1,2,…$x$ refer to the number of nanorods in the chain.

By inserting into Eq. 5 the extinction cross-sections obtained from the FDTD simulations and $c_{x,p}$ obtained from Eq. 4, we calculated the extinction spectra of the entire population of nanorod chains at various values of $p$ (corresponding to particular self-assembly times $t$). Extinction spectra for $p$ of 0.3, 0.5, 0.7 and 0.9 are shown in Fig. 2(b-e), respectively. The thin lines show the contribution to the overall extinction spectrum by the nanorod chains with a particular $x$, that is, $b\sigma_x(\lambda)c_{x,p}$, and the thick lines show the total extinction of all the chains ($Ext_{tot,p}(\lambda)$).

**Monte-Carlo modeling of polydisperse nanorod chains**

In an experimental setting, synthesized gold nanorods always exhibit polydispersity. In the present work, we assume that the nanorods have lengths and diameters with distributions of $L$ = 52±6.1 and $d$ = 13±1.6 nm, respectively, and the distance between the ends of nanorods in the self-assembled chains is $l$ = 6.7±1.4 nm (see Appendix for transmission electron microscopy (TEM) images from which these values are inferred (Fig. A1)). Empirically (and as a consequence of the central limit theorem), these distributions are approximately Gaussian.

To model the extinction spectrum of a collection of nanorod chains with geometrical variations, we could in principle perform an exhaustive set of FDTD simulations, sweeping over all possible nanorod lengths and diameters, as well as over all of the possible gap lengths between adjacent nanorods in the chains, and weigh the resulting spectra appropriately to predict the expected LSPR spectrum of the ensemble (in the Appendix, we show this type of calculation for nanorods with just one polydisperse parameter using a semi-analytical approach). However, even for a modest number of nanorod constituents of each chain, this parameter space explodes, making this computational problem intractable. To overcome this, we employed a strategy which combines the Monte Carlo method with FDTD simulations, which is graphically described in Fig. 3. For a chain comprising $x$ nanorods, we assumed that the geometrical parameters $L$, $d$, and $l$, for each nanorod and gap are independent and identically Gaussian distributed throughout the chain. Accordingly, we selected the geometrical parameters for each nanorod and each gap stochastically from the appropriate empirical Gaussian distribution (Fig. 3(a)), and then used FDTD simulations to calculate the normalized LSPR spectrum for the nanorod chain (Fig. 3(b)). We iterate this process until a relatively smooth, invariant distribution emerges from the average of the simulated spectra, and then fit this average spectrum to a Gaussian distribution to obtain an estimate of the average extinction spectrum $\sigma_x'(\lambda)$ (Fig. 3(c)). For this work, we performed 250 simulations for the monomers, 150 simulations for the dimers, 90 for the 3-mers, and 60 each for chains comprising 4-10 nanorods. While the results could be made more accurate by utilizing more simulations, this number of simulations was enough to demonstrate the effects of inhomogeneous broadening, and was a compromise between accuracy and computational resources. The resulting Gaussian fits to the normalized extinction spectra are shown in Fig. 4(a).

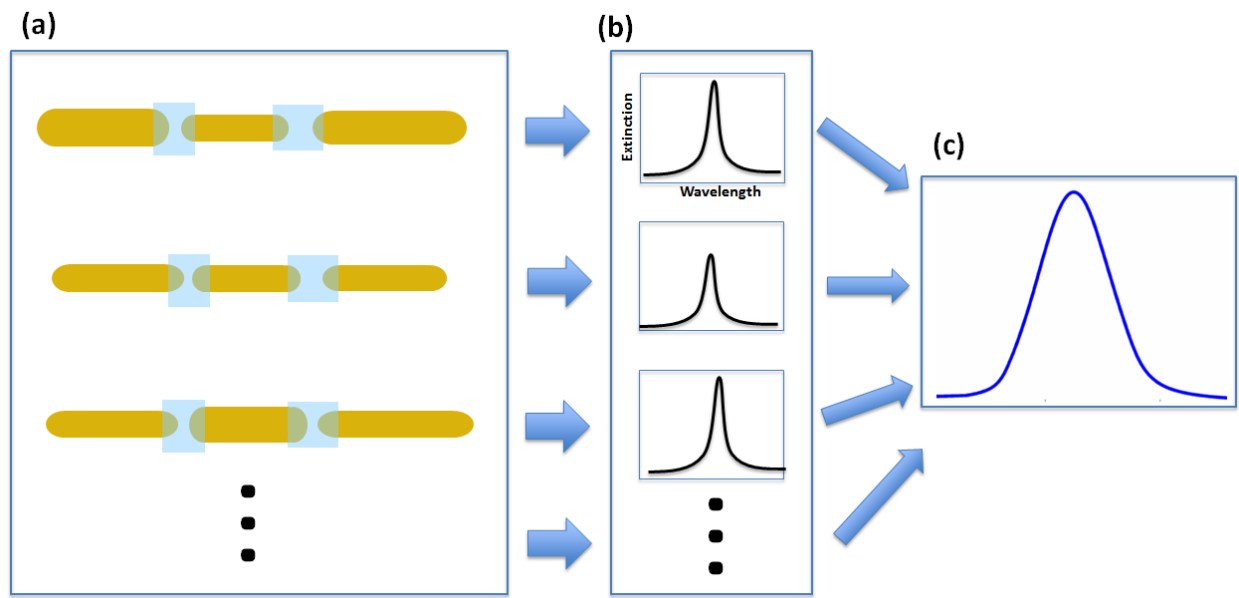

Figure 3. Schematic describing the Monte-Carlo technique for calculating the extinction spectrum of a collection of trimer ($x = 3$) chains comprising polydisperse nanorods. (a) First a number of chains are stochastically generated, with each nanorod length ($L$), width ($d$), and gap length ($l$) selected from an empirically-determined Gaussian distribution. (b) FDTD simulations are used to determine the extinction spectra of each chain generated in (a). (c) The simulated spectra from (b) are averaged and fitted to a Gaussian to obtain a predicted spectrum of a collection of polydisperse trimers.

Note that we expect this spectrum of an ensemble of polydisperse nanorod chains to resemble a Gaussian moreso than a Lorentzian distribution (as would be expected for a single, isolated resonance) as a consequence of a general correlation between resonance peak wavelengths and the overall lengths of their corresponding nanorod chains [39]. Since the nanorod chain lengths are Gaussian-distributed, the resonance peak wavelengths tend to be Gaussian-distributed as well. Since the widths of these Gaussian distributions of resonance peaks tend to be greater than the widths of the individual resonances, we expect that the ensemble spectrums will have more Gaussian than Lorentzian character. This is verified for the special case of unreacted polydisperse nanorods in the Appendix (Fig. A1).

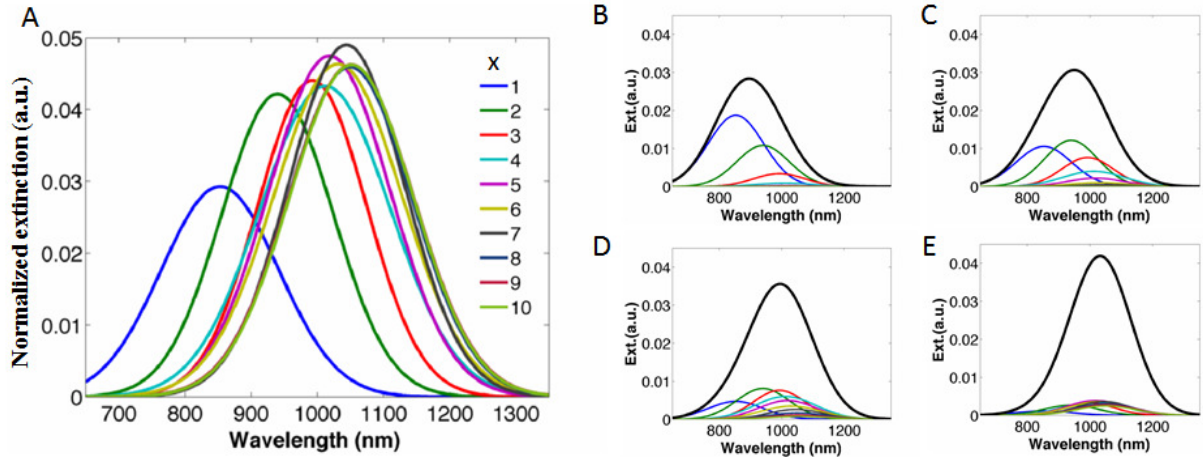

Figure 4. Simulated extinction spectra of chains comprising polydisperse nanorods. (a) Gaussian fits to the normalized extinction spectra of chains for $x$ from 1 to 10 as calculated by our FDTD-based Monte Carlo method. The lengths and widths of each nanorod, and the gap lengths between them, were stochastically selected from Gaussian distributions where $L = 52\pm6.1$ nm, $d = 13\pm1.6$ nm, and $l = 6.7\pm1.4$ nm (b-e) Simulated extinction spectra of ensembles of polydisperse nanorod chains with conversions $p$ of 0.3, 0.5, 0.7, and 0.9, respectively. The thin colored lines show the contribution to the overall extinction spectrum from the chains with a particular $x$, and the thick black lines show the total extinction spectrum of the entire population of chains ($Ext_{tot,p}$).

As $x$ increased from 1 to 10 we again see red-shift of the LSPR peak from ~850 nm to ~1050 nm (Fig. 4(a)). The normalized extinction spectra of individual chains comprising polydisperse nanorods are substantially broader than their monodisperse counterparts. This broadening is not a result of increased absorption or scattering losses which would also lead to a broader peak in extinction due to decreased quality factors, but is instead a result of inhomogeneous broadening. While it is evident from Fig. 4(a) that the normalized LSPR peak heights and locations have not fully stabilized (more simulations would be necessary for the results to fully converge), we can still clearly see the trends in normalized LSPR extinction spectra as $x$ increases. For example, while in monodisperse chains the normalized LSPR peak heights tend to decrease slightly as $x$ increases (Fig. 2(a)), the opposite is true for polydisperse chains. This is because the polydispersity has a stronger effect on the chains comprising fewer nanorods than on the longer chains: in the longer chains the small variations in the individual rods tend to cancel out; this means that the normalized spectra of the ensemble of shorter polydisperse chains tend to be broader and correspondingly smaller in amplitude compared to ensembles of longer chains. In this way, the Monte Carlo approach yields a qualitatively different prediction than that obtained from a monodisperse approximation.

By inserting the averaged and fitted values of $\varepsilon_x(\lambda)'$ (obtained from the Monte Carlo simulations) and $c_{x,p}$ (obtained from Eq. 4 according to the Flory distribution) into Eqn. 5, we calculated the extinction spectra of the entire population of polydisperse nanorod chains at various values of conversion $p$. Extinction spectra for $p$ values of 0.3, 0.5, 0.7 and 0.9 are shown in Fig. 4(b-e), respectively, along with the contributions of individual populations of nanorod chains of various $x$. The total extinction spectrum of all chains for $p$ ranging from 0 to 0.9 is shown in Fig. 5(b), compared to the same calculation using monodisperse chains shown in Fig. 5(a). In the monodisperse case, the total extinction spectra of intermediate $p$ values are distinctly bimodal due to the relatively narrow LSPR peak widths associated with individual nanorod chains and the relatively large LSPR peak red-shifts associated with increasing $x$. In the polydisperse case, the relatively broad LSPR peak widths which result from inhomogeneous broadening wash out much of this bimodal spectral feature. Instead, the polydisperse spectra each feature a single broad peak which slowly red-shifts with increasing self-assembly time (and hence conversion $p$).

In this paper we intentionally make no comparison to experimental optical data. While our simulation method effectively accounts for many key physical effects contributing to the optical response of self-assembled nanopolymer solutions including the polydispersity of nanorod widths and lengths and gaps between nanorods, it does not account for chain bending, retardation effects for chains which are not oriented perpendicular to the incident light, or additional broadening effects from thermal motion of the nanorod chains. Our model system of self-assembled nanopolymers in solution is a particularly challenging system to simulate due to the overall number of degrees of freedom and the computational resources required for every full-wave 3D simulation of large nanorod chains with small features (such as gap sizes) which must be well resolved. Despite this, we believe that the Monte-Carlo approach combined with electromagnetic simulations (or analytical calculations) which we demonstrate here may be the most efficient method that provides meaningful information about inhomogeneous broadening in optical systems, especially when applied to slightly simpler systems such as lithographically-defined or self-assembled nanostructures on a substrate.

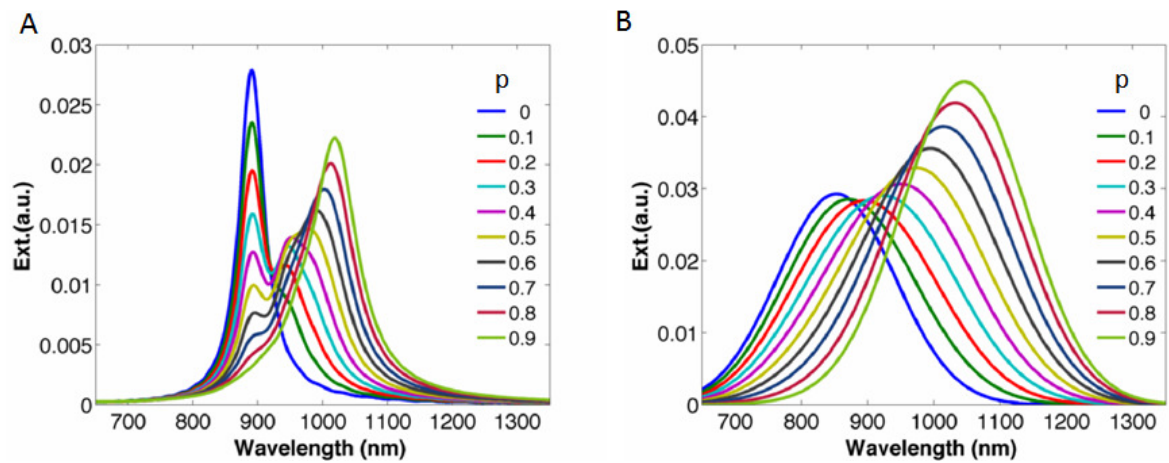

Figure 5. Summary of the simulated extinction spectra of the Flory-distributed nanorod chains for conversions *p* from 0 (all unreacted single nanorods) to 0.9 (polymerization process nearly complete). In (a), the nanorods comprising the chains are monodisperse, whereas in (b) the nanorod widths, lengths, and gap lengths are each normally distributed around the same mean used in (a).

**Conclusion**

In conclusion, we utilized finite difference time domain (FDTD) simulations combined with a Monte Carlo approach to study the effects of inhomogeneous broadening on the extinction spectra of populations of gold nanorods formed in solution. We found that inhomogeneous broadening due to dispersion in the geometrical parameters of the nanorods (lengths and widths) and the gaps between neighboring rods significantly affected the shape and bandwidth of the resonance spectra of solutions of nanorod chains. More generally, we conclude that in systems involving large collections of independent resonant elements, inhomogeneous broadening introduces significant differences between the resonant responses of individual elements and the ensemble. To account for such differences, it is possible to run separate calculations or simulations for every possible set of geometrical parameters and then perform a weighted average; however, as in the present demonstration, this is often an intractable problem, especially for structures with many degrees of freedom and resource-expensive numerical techniques such as FDTD. This, however, can be overcome by using a Monte Carlo approach consisting of iterated stochastic sampling from the entire parameter space combined with numerical simulations. In the present demonstration, three-dimensional full-wave FDTD simulations are used, but in principle any analytical, semi-analytical, or fully numerical method can be applied.

**Acknowledgements**

This work was supported in part by the Defense Advanced Research Projects Agency (DARPA) N/MEMS S&T Fundamentals program under grant no. N66001-10-1-4008 issued by the Space and Naval Warfare Systems Center Pacific (SPAWAR). E. K. thanks NSERC Canada for supporting this work by a Discovery Grant and Strategic Network for Bioplasmonic Systems Biopsys Grant. The Lumerical FDTD simulations in this article were run on the Odyssey cluster supported by the Harvard Faculty of Arts and Sciences Division Research Computing Group. MK acknowledges helpful discussions with N. Yu, R. Blanchard, and J. Fan and is supported by the National Science Foundation through a Graduate Research Fellowship.

## Supplementary Information

Semi-analytical calculation of single nanorods

For the case of nanorod monomers, in which there are only two degrees of freedom (length and width), we can semi-analytically calculate the expected average extinction spectrum of a polydisperse ensemble. As described by Prescott and Mulvaney, the extinction spectrum of a uniform ellipsoid dispersed in a nonabsorbing medium can be described by its dielectric function and appropriate geometrical factors [1]. Using the geometrical factors for cylindrical nanorods with spherical ends, we computationally averaged the extinction spectra of gold nanorods with $40 \pm 4.7$ nm Gaussian distributed lengths, keeping the diameter $d$ constant at 10 nm for simplicity, sampling 188 lengths across two standard deviations (Fig. A1). These values do not exactly correspond to the values used in the main text, and are used for illustrative purposes only. In Fig. A1(a), we plotted twenty individual extinction spectra for increasing nanorod length, and in Fig. A1(b) we plotted the average of all 188 spectra (blue line). We attempted to fit this data to Gaussian and Lorentzian lineshapes in wavelength, and observed that the averaged extinction spectrum is substantially more Gaussian than Lorentzian in character. The Gaussian fit is a very good approximation at or around the resonance, though it falls off quicker than the actual weighted average of the extinction spectra.

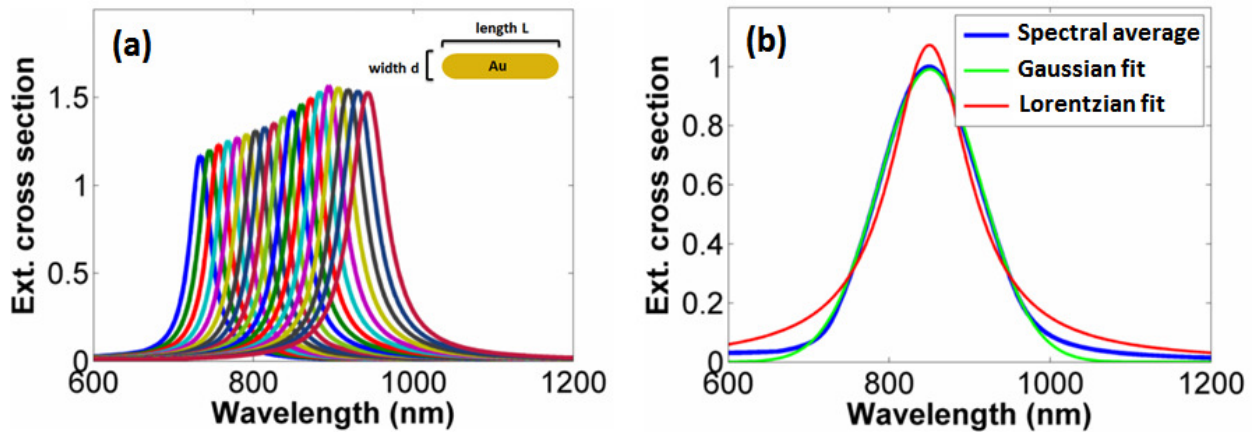

Figure A1. (a) Extinction cross-sections of gold nanorods with diameter $d = 10$ nm, and length $L$ varied from 30.6 nm to 49.4 nm. Inset: schematic of a nanorod of length $L$ and diameter $d$. (b) Average of the spectra in (a), weighted by a Gaussian distribution in length centered around $L = 40$ nm with a standard deviation of 4.7 nm (blue). Green and red curves are the Gaussian and Lorentzian fits, respectively.

Convergence

Evaluating the convergence of a Monte Carlo experiment is generally a challenging task. Sometimes the rate of convergence of the Monte Carlo sample mean to the true mean can be estimated using the Monte Carlo sample variance by appealing to the central limit theorem [2]. Our case is somewhat complicated by the fact that we are examining the convergence of extinction spectra. While one could estimate convergence criterion based on point-by-point sample variances in the extinction spectra, in our case it is more appropriate to evaluate convergence of the Gaussian fits to the spectra. In Fig. A2, we see that the relative fluctuations in Gaussian fit means and standard deviations of individual nanorod chains generally decrease with increasing sample size. In particular, monomer, dimer, and trimer chains required substantially more Monte Carlo samples to achieve degrees of stability comparable to that of longer chains. We found that the extinction spectra of individual Monte Carlo samples for shorter chains were substantially less clustered than those for longer chains. This highlights the fact that the convergence of Monte Carlo experiments is greatly situational, and must be evaluated case-by-case. While additional sampling is required for the solution to truly converge (see Fig. 4(a) and Fig. A2(a, b)), the large computational cost associated with achieving substantially finer convergence deterred us from additional sampling.

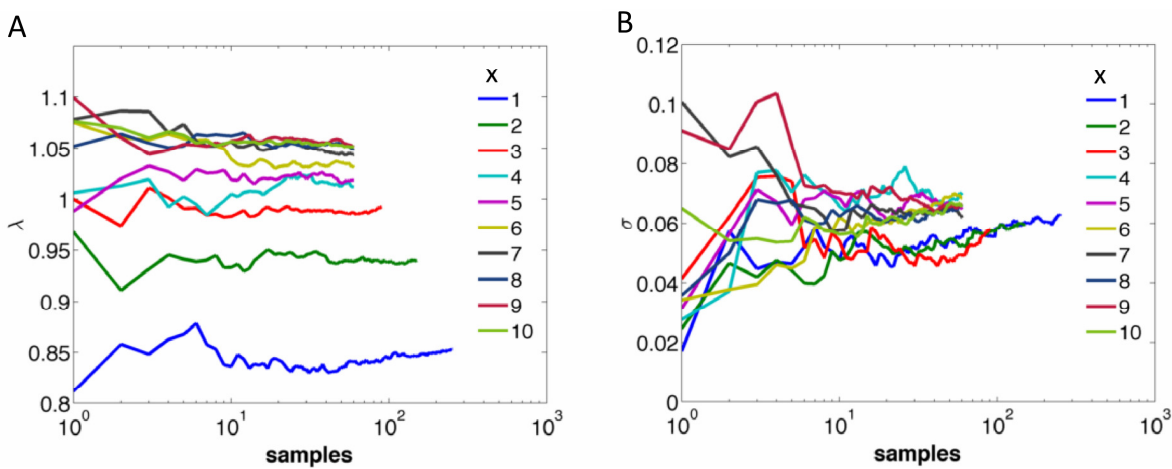

Figure A2. Convergence of average Monte Carlo based simulations. (a) mean of Gaussian fits of average extinction spectra as sample number is increased. (b) standard deviation as in (a).

Geometrical parameters for nanorods used in simulations

The lengths and widths of each nanorod, and the gap lengths between them, were stochastically selected from Gaussian distributions where $L = 52\pm6.1$ nm, $d = 13\pm1.6$ nm, and $l = 6.7\pm1.4$ nm

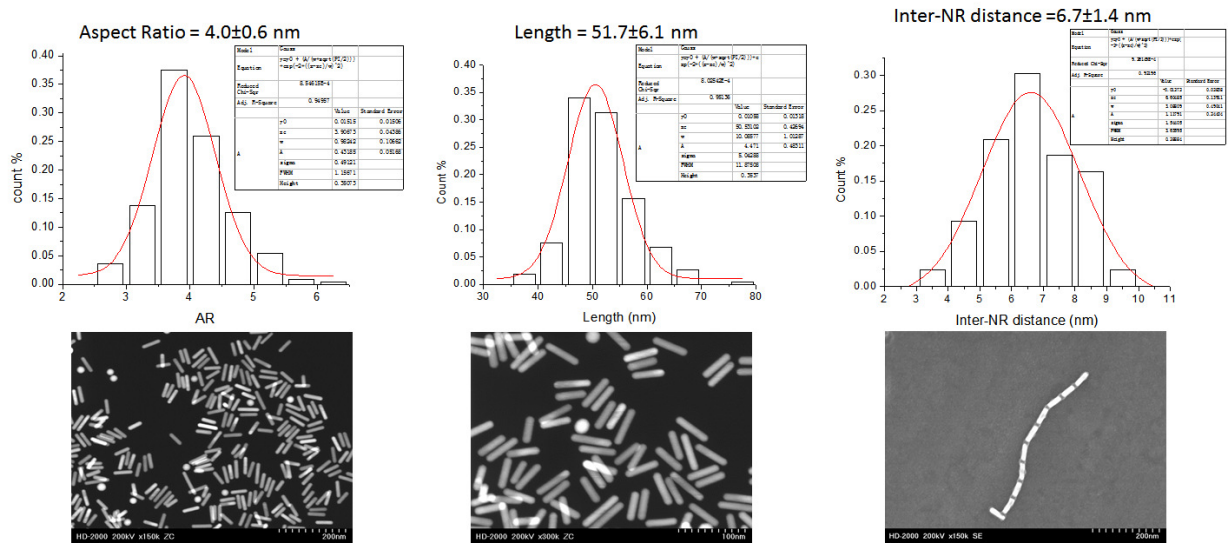

Figure A3. Transmission electron microscope (TEM) images of gold nanorods and nanorod chains in solution. We performed statistical analysis on the lengths *L*, aspect ratios *d/L*, and inter-nanorod distances *l* to obtain the distributions used in the calculations presented in the main text.

**Supplementary References**

[1] Prescott SW, Mulvaney P (2006), Gold nanorod extinction spectra, *Journal of Applied Physics* 99(12):123504

[2] Ata MY (2007), A convergence criterion for the Monte Carlo estimates, *Simulation Modeling Practice and Theory* 15(3):237-246